\documentclass[aps,preprint,pre]{revtex4}
\usepackage{epsfig}
\usepackage{graphicx}
\begin{document}
\title{Above-well, Stark, and potential-barrier resonances of an open square
well in a static external
electric field}
\author{A. Emmanouilidou$^{1}$, N. Moiseyev$^{2}$}
\affiliation{$^{1}$Max Planck Institute for the Physics of Complex Systems,
N$\ddot{o}$thnitzer Stra$\beta$e 38, 01187 Dresden, Germany }
\affiliation{$^{2}$ Department of Chemistry and Minerva Center of Nonlinear
Physics in Complex Systems Technion-Israel Institute of Technology Haifa
32000, Israel}
\date{\today}

\begin{abstract}
Besides the well known Stark resonances, which are localized in the potential well
and tunnel through the potential barrier created by the dc-field,
 ``strange" long and short-lived resonances are analytically obtained.
These resonances are not localized inside the potential well.
 We show that the narrow ones are localized above the potential well. These
narrow resonances give rise to a {\it peak structure} in a 1D
scattering experiment.
 We also show that the broad overlapping resonances are associated with the
static electric field potential barrier.
 These ``strange" overlapping resonances do not give rise to a {\it peak structure}
in a 1D scattering experiment. We propose a 2D experimental set-up
where in principle these short-lived states should be observed as
{\it peaks}.
 Broad overlapping resonances,
 associated only with the static electric field potential barrier, could also have observable effects in a $N>1$ array of
quantum wells in the presence of a truncated static electric field. This last problem is associated with the resonance tunnelling
phenomena which are used in the construction of resonance-tunnelling diodes
and transistors.

\end{abstract}
\maketitle

PACS number(s): 32.60.+i, 34.50.-s
\section{Introduction}
 Elucidating the mechanisms underlying the decay processes in
mesoscopic as well as atomic systems is of fundamental importance.
 Valuable insight into the decay mechanisms of these systems is obtained
when one expresses their dynamical properties in terms of the resonance
states of the system (complex eigenvalues in the lower half energy plane).
 Resonances in the presence
of a static electric field that are directly associated with the bound or
resonance states of the field-free Hamiltonian, known as Stark resonances, are
well understood and studied for many years \cite{Reinhardt}. Less known and understood are resonances in the presence
of a static electric field that can not be traced back to the bound, excited or resonance states of the field-free Hamiltonian.

 A simple model where the later type of resonances appear is
the one-dimensional $\delta$-potential well in the presence of a static
electric field. For this simple model, Ludvikson has
shown the existence of two families of an infinite number of resonances that can
not be traced back to the bound state of the field-free Hamiltonian \cite{Ludvikson}. The effect of these families
of resonance states in physically measurable quantities was studied by
Emmanouilidou and Reichl \cite{Agapi}. In this latter study, the authors, using a complex spectral decomposition \cite{Nickel} for
Ludvikson's one-dimensional model, were able to qualitatively reproduce
 the main features of the experimentally obtained photodetachment
cross section of the $H^{-}$ ion in the presence of a static electric field when driven
by a weak time-periodic electric field \cite{Hion1, Hion2} and directly associate these features
to the two-families of infinite resonance states. More recently, Ludvikson's
model was studied in ref\cite{Alvarez}, and using two-$\delta$ potentials in ref\cite{Korsch} in the context of avoided crossings.

 Open questions yet remain regarding the ``nature" of these resonance
states which do not stem from the spectrum of the field-free
Hamiltonian. In the current work we address some of these
questions using as our model system a one-dimensional square well
in the presence of a static electric field. We show that this
model potential supports Stark resonances that are localized
inside the square well and tunnel through the potential barrier
created by the dc-field. We also find two families of infinite
resonances that can not be traced to the spectrum of the
field-free Hamiltonian. One family consists of long-lived
resonance states that are localized above the potential well,
while the other family consists of short-lived overlapping
resonances that are localized below the potential well. Resonance
states similar to the long-lived ones that are not associated with
the field-free Hamiltonian have been probed as {\it peaks} in
experiments measuring the photodetachment rate of negative ions
driven by a weak time periodic field in the presence of a static
field \cite{Hion1,Hion2}. However, little is known about the
``nature" of the short-lived states that are not associated with
the field-free Hamiltonian. Since these resonance states do not
give rise to a {\it peak structure} in a 1D scattering experiment,
it is an open question what is an appropriate experimental set up
where these short-lived overlapping resonances could be probed as
{\it peaks}.

  We shed light into the nature of these
resonance states, by showing first that they are associated {\it only
with the static electric field potential barrier}.
 Next, by introducing a cut-off to the static electric field we find
 overlapping resonances
 similar to the broad overlapping resonances that are present
 in the case of the square well plus static electric field.
  In both cases, with and without the cut-off, the square well plus static
electric field supports overlapping resonances that are {\it only associated
with the static electric field potential barrier}.
  Using exterior complex scaling, we show that the
  overlapping resonances of the truncated potential have a nodal
  structure.
 Future work might involve studying the effect of the long and short-lived resonances on the
 transmission properties through an
$N>1$ array of 1D quantum wells in the presence of a static electric field. This latter
problem is associated with the resonance tunnelling phenomena which are used
in the construction of resonance-tunnelling diodes and transistors. Our results
can be perhaps used
in the design of a new type of controlled electronic switches.
 Finally, we also propose a two-dimensional scattering set-up where in principle
the short-lived states of the square well plus static electric field should
  give rise to a {\it peak structure}. The
proposed 2D set-up is based on an idea proposed by Narevicius and
Moiseyev\cite{Narevicius} involving a light atom getting trapped
among two heavy ones.

\section{Model}
The model we use describes the behaviour of a single particle, of mass $m$
and charge $q$, in one space dimension, in the presence of a static electric field
and a square well potential. The one dimensional Hamiltonian is:
\begin{equation}
H(\xi)=-\frac{\hbar^{2}}{2m}\frac{\partial^{2}}{\partial \xi^{2}}-F\xi-V(\xi),
\label{eq:Hamiltonian1}
\end{equation}
where $V(\xi)$ is given by:
\begin{equation}
\label{potential}
V(\xi)=\left\{\begin{array}{lll}
 -\xi, & \mbox{$\xi< -\alpha'$} \\
 -V'_{0}-\xi,& \mbox{$-\alpha'<\xi< \alpha'$}\\
  -\xi, &\mbox{$\xi>\alpha'$}
                             \end{array}
                              \right.
\end{equation}
with $F/q$ the strength of the static electric field and $V'_{0}$ the depth
of the square well. If we introduce the dimensionless variables
\begin{equation}
\label{eq:dim}
\begin{array}{cccccc}
x=\xi/\xi_{0},   & E=E'/\epsilon_{0}, &   t=t'\epsilon_{0}/\hbar,
& V=V'/\xi_{0}\epsilon_{0}, &z=z'/\epsilon_{0},
\end{array}
\end{equation}
where
\begin{equation}
\label{eq:dim1}
\begin{array}{cc}
\xi_{0}=(\frac{\hbar^2}{2mF})^{1/3},& \epsilon_{0}=F\xi_{0}
\end{array}
\end{equation}
the Hamiltonian is given by
\begin{equation}
\label{Hamiltonian2}
H(x)=-\frac{\partial^{2}}{\partial x^{2}}-x-V(x)
\end{equation}
with
\begin{equation}
\label{eq:potential1}
V(x)=\left\{\begin{array}{lll}
 -x, & \mbox{$x< -\alpha$} \\
 -V_{0}-x,& \mbox{$-\alpha<x< \alpha$}\\
  -x, &\mbox{$x>\alpha$}
                             \end{array}
                              \right.
\end{equation}
Note that $t'$, $z'$ are the time, and the energy on the complex plane variables, respectively,
and $t$, $z$ are the corresponding dimensionless variables. 
 The above Hamiltonian has a continuous energy
spectrum in the Hilbert space. One way to obtain the resonance states of the
system, which are given by complex eigenvalues in the lower energy plane, is
to obtain the poles of the analytically continued energy Green's function in
the lower half energy plane.
 In the current work, we employ a different approach that allows us to find the
resonance states of the system in a {\it very simple way}. In this simple approach, the wavefunction
satisfies Gamov-Siegert boundary conditions, that is, it satisfies
outgoing wave boundary conditions:

\begin{equation}
\label{eq:wave1}
\Psi (x, z)=\left\{\begin{array}{lll}
 N Ai(-x-z), & \mbox{$x< -\alpha$} \\
 N \left(a_{1}(z)Ai(-x-V_{0}-z)+a_{2}(z)Bi(-x-V_{0}-z)\right),& \mbox{$-\alpha<x< \alpha$}\\
 N a_{3}(z)Ci^{+}(-x-z), &\mbox{$x> \alpha$}
                             \end{array}
                              \right.
\end{equation}
where $Ci^{+}$ describes an outgoing wave in the positive $x$
direction in the presence of a static electric field and $Ci^{+}=iAi+Bi$. For
$z$ in the lower half energy plane, the wavefunction $\Psi(x,z)$ goes
to zero as $x\rightarrow -\infty$ but is not bounded as
$x\rightarrow\infty$. Thus, due to the outgoing wave boundary conditions we
imposed, $\Psi(x,z)$ is not a square integrable
wavefunction. From the continuity of the wavefunction and its first
derivative at $x=-\alpha$ and $x=\alpha$ we find $a_{1}(z)$, $a_{2}(z)$,
 $a_{3}(z)$:
$$a_{1}=\pi(Ai(\alpha-z)Bi^{'}(\alpha-V_{0}-z)-Ai^{'}(\alpha-z)Bi(\alpha-V_{0}-z),$$

$$a_{2}=\pi(Ai^{'}(\alpha-z)Ai(\alpha-V_{0}-z)-Ai(\alpha-z)Ai^{'}(\alpha-V_{0}-z)),$$

$$a_{3}=(a_{1}Ai(-\alpha-V_{0}-z)+a_{2}Bi(-\alpha-V_{0}-z))/Ci^{+}(-\alpha-z),$$
as well as the condition for obtaining the resonances, $z_{n}$, of the system:

\begin{equation}
\label{eq:condresonances}
\begin{array}{llllll}
(Ai(-\alpha-V_{0}-z_{n})iAi'(-\alpha-z_{n})+Ai(-\alpha-V_{0}-z_{n})Bi'(-\alpha-z_{n})- \\
Ai'(-\alpha-V_{0}-z_{n})iAi(-\alpha-z_{n})-Ai'(-\alpha-V_{0}-z_{n})Bi(-\alpha-z_{n}))\\
(Ai(-\alpha-z_{n})Bi'(\alpha-V_{0}-z_{n})-Ai'(\alpha-z_{n})Bi(\alpha-V_{0}-z_{n})+\\
(Bi(-\alpha-V_{0}-z_{n})Bi'(-\alpha-z_{n})+Bi(-\alpha-V_{0}-z_{n})iAi'(-\alpha-z_{n})-\\
Bi'(-\alpha-V_{0}-z_{n})iAi(-\alpha-z_{n})-Bi'(-\alpha-V_{0}-z_{n})Bi(-\alpha-z_{n}))\\
(Ai'(\alpha-z_{n})Ai(\alpha-V_{0}-z_{n})-Ai(\alpha-z_{n})Ai'(\alpha-V_{0}-z_{n}))=0.
\end{array}
\end{equation}
Using Eq.(\ref{eq:condresonances}) we find the resonances of the square well
plus static electric field for $V_{0}=5$ and $\alpha=1$, shown in Table I, and their
corresponding wavefunctions satisfy:
\begin{equation}
\label{eq:Hilbert}
H(x)\Psi(x,z_{n})=z_{n}\Psi(x,z_{n})
\end{equation}
with $\Psi(x,z_{n})$ not an element of the Hilbert space and $H(x)$ being non Hermitian.

\section{Stark resonances}

The square well plus static electric field supports resonance states
that stem from the bound as well as the resonance states of the field-free
square well. In Table I, the resonances denoted by $0$ and $1$ stem from
 the bound states of the field-free square well and in what follows we refer to
them as type I Stark resonances. Indeed, one can easily show that the conditions for bound states of the
square well in dimensionless units are:
$$-\sqrt{E}+\sqrt{V_{0}-E}tan(\sqrt{V_{0}-E})\alpha)=0$$
$$\sqrt{E}+\sqrt{V_{0}-E}cot(\sqrt{V_{0}-E})\alpha)=0.$$
From the above conditions one finds that for $V_{0}=5$ the square well supports two bound states, one with even
symmetry and value $E=-3.8525$ and one with odd symmetry and value
$E=-0.9314$. These energies are very close to the real part
of the resonances $0$ and $1$, respectively, in Table I.

 In addition to the type I Stark resonances, the square well plus static
electric field
supports resonance states (type II Stark resonances) that stem from the
resonance states of the field-free square well. When computing the
transmission probability for the one-dimensional field-free square well one
finds {\it peaks} for energies equal to the real part of the resonance states. The field-free square well
supports resonance states \cite{Schwabl} whose real part is given by:
\begin{equation}
\label{eq:squareresonances}
E_{R,l}=(\frac{l\pi}{2\alpha})^{2}-V_{0},
\end{equation}
where $(\frac{l\pi}{2\alpha})^{2}$ with $l=1,2,...$ are the energy levels of an infinite square
well that extends from $-\alpha$ to $\alpha$.
Using Eq.(\ref{eq:squareresonances}), we find that for $V_{0}=5$ the
lowest resonance state of the field-free square well has a real part that is given by
$E_{R,2}=4.8696$. Note, that this energy is close to the real part of
the resonance state denoted by $n=3$ in Table I.

\begin{figure}
\begin{centering}
\leavevmode
\epsfxsize=0.6\linewidth
\epsfbox{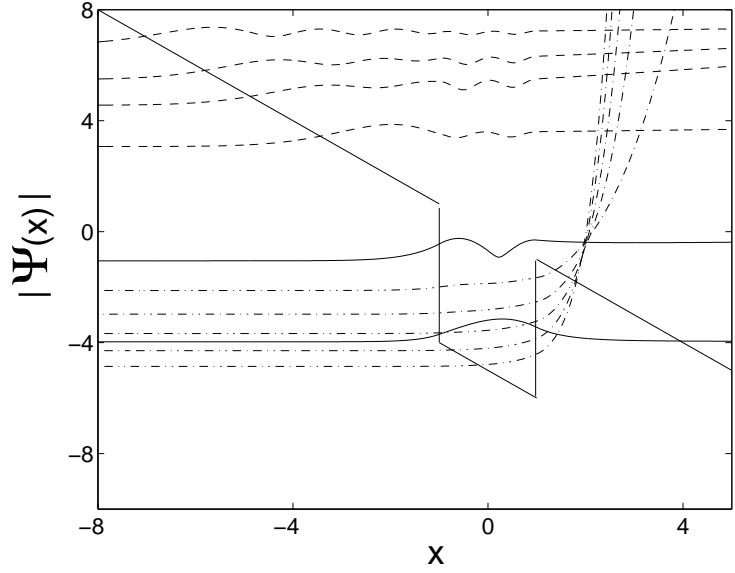}
\caption{Absolute value of the wavefunction $|\Psi_{n}(x)|$. The plot for each
wavefunction is shifted by the real part of the resonance state $z_{n}$. The
solid lines indicate the type I Stark resonances, the dashed lines indicate
the long-lived ``new-born" resonance states, while the dashed-dot lines
indicate the short-lived ``new-born" resonance states.
 The normalization constant in Eq.(\ref{eq:wave1}) is evaluated so that
$\int_{-\infty}^{-\alpha}\Psi^{2}(x,z_{n})dx+\int_{-\alpha}^{\alpha}\Psi^{2}(x,z_{n})dx-[\int\Psi^{2}(x,z_{n})dx|]_{x=\alpha}=1$.}
\label{fig:absvalue}
\end{centering}
\end{figure}

 In Fig.(\ref{fig:absvalue}), we plot the absolute value of the resonance
 wavefunctions $\Psi_{n}(x)$ as a function of the one-dimension $x$.
 The normalization constant in Eq.(\ref{eq:wave1}) is evaluated so that
 \begin{equation}
 \label{eq:norm10}
\int_{-\infty}^{-\alpha}\Psi^{2}(x,z_{n})dx+\int_{-\alpha}^{\alpha}\Psi^{2}(x,z_{n})dx-[\int
 \Psi^{2}(x,z_{n})dx]|_{x=\alpha}=1.
\end{equation}
 Using this definition of norm the wavefunctions corresponding to different
 resonances are orthogonal, for more details on using this norm see
 ref\cite{Zeldovich}. The
Stark resonances of the square well plus static electric field are localized inside the
 potential well and exhibit a nodal structure. Note that in order for one to 
see the antinode-node structure one would have to plot separately the real and
 the imaginary part of the wavefunction. In the current work we plot the
 absolute value of the wavefunction and that is the reason why only the maxima
 of the wavefunction are exposed. For brevity we refer to this structure as
 nodal. This can be clearly seen for
 the type I Stark resonances that are plotted in
 Fig.(\ref{fig:absvalue}). Less clear is the nodal pattern of the type II
 Stark resonance $z_{3}$ due to its interference
 with the long-lived resonance states stemming from
 the continuum that are localized above the barrier, as we discuss in the following
 section. A comparison with Fig.(\ref{fig:V030})
reveals that the resonance state $z_{3}$ has a nodal pattern with two maxima
inside the square well since $z_{3}$ seems to be associated with the $E_{R,2}$
resonance state of the field-free square well. A more detailed study of the
resonance dependence with the strength of the static electric field would
 determine beyond doubt the resonance states in Table I
  that are associated
with the resonances in the field-free square well.

\section{``New-born" resonance states that are not associated with
the poles of the field-free potential well.}

In addition to the Stark resonances, the square well plus static
electric field supports resonance states that can not be traced back to the
bound or resonance state poles of the field-free square well. We
refer to these resonance states as ``new-born" states.
 There are two families of an infinite
number of these resonance states with one family consisting of
long-lived states and the other one consisting of short-lived
ones.

\subsection{ Long-lived ``new-born" above-well resonance states}

For $V_{0}=5$ and $\alpha=1$, the family of long-lived states stemming from
the continuum are the $z_{2}, z_{4}, z_{5},...$ resonances
in Table I. Using Eq.(\ref{eq:condresonances}), one can show that when $V_{0}\rightarrow \infty$ the family of the long-lived resonance states
stemming from the continuum are given by $Ai(\alpha-E)=0$ \cite {Ludvikson}.
 That is, to the lowest order in 1/V,
 the real part of these long-lived states is given by
$\alpha+(-3\pi/8+3l\pi/2)^{2/3}$, with $l=1,2,...$. In Table II we show the real part of the
first eight long-lived states when $V_{0} \rightarrow \infty$.
 The value of $V_{0}=5$ and $\alpha=1$ we use to find the resonance states
corresponds to a deep potential well. A comparison of Table I and Table II
reveals that the real part of the resonance states $z_n$ with $n\ge 6$
is very similar to the energies in Table II with $l>4$. The resonance
states $z_{2}, z_{4}$ and $z_{5}$, in Table I, have a real part not as close to the
energies corresponding to $l=1,2$ and $l=3$ in Table II, due to the interference
of the family of the long-lived resonance states with the $z_{3}$ Stark resonance.

 When the
square well becomes very deep the family of the long-lived
resonance states are very close to the eigenstates of the
triangular well that is formed by the static electric field in the region $x<-\alpha$
and a wall at $x=-\alpha$. In Fig.(\ref{fig:absvalue}) we see that the long-lived states, denoted by the dashed curves in Fig.(\ref{fig:absvalue}),
 are localized above the barrier and have a nodal structure inside
the triangular well. Due to the interference with the $z_{3}$
Stark resonance, the nodal pattern of the family of long-lived
states has also a nodal structure consisting of two maxima in the
region where the square well extends. Long-lived ``new born" resonance states
 similar to the ones currently under consideration have been known to give
rise to a {\it peak
 structure} in photodetachment experiments of negative ions driven by a weak-time
periodic field in the presence of a static electric field \cite{Hion1,Hion2}.

\begin{figure}
\begin{centering}
\leavevmode
\epsfxsize=0.6\linewidth
\epsfbox{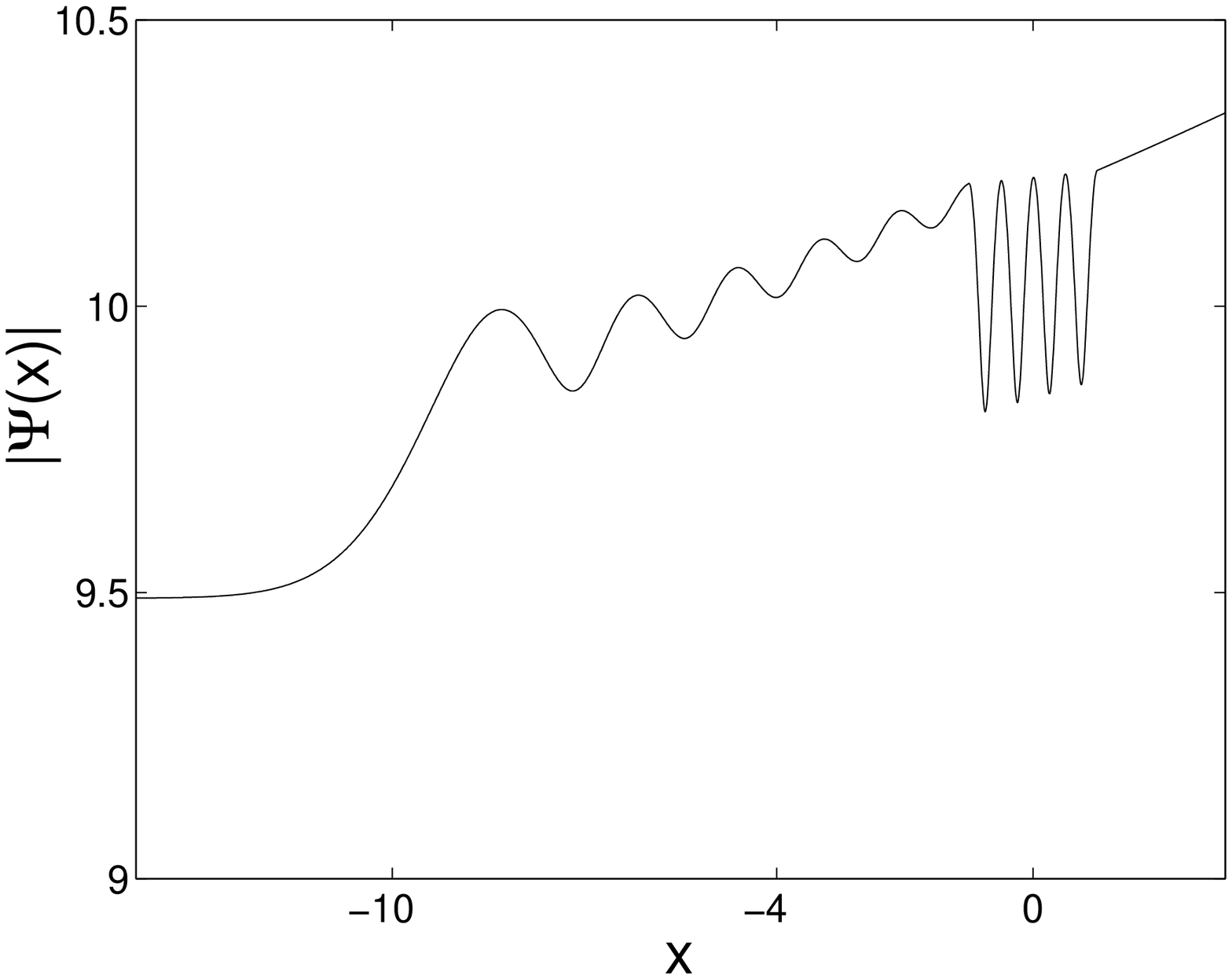}
\caption{Absolute value of the wavefunction $|\Psi_{z_{6}}(x)|$. The plot is shifted by the real part of the resonance state $z_{6}$.}
\label{fig:V030}
\end{centering}
\end{figure}

\begin{figure}
\begin{centering}
\leavevmode
\epsfxsize=0.6\linewidth
\epsfbox{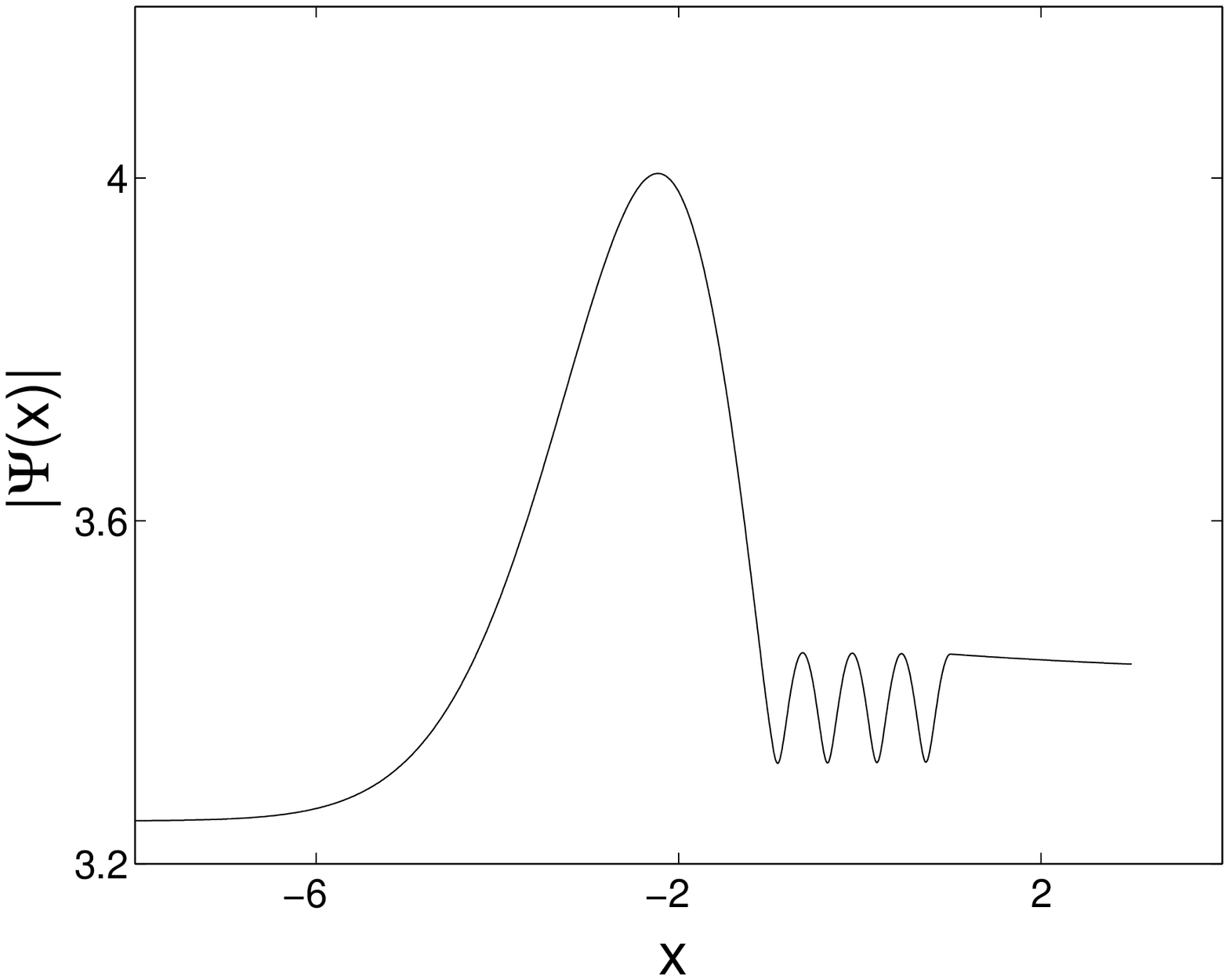}
\caption{Absolute value of the wavefunction $|\Psi_{z_{1}}(x)|$. The plot is shifted by the real part of the resonance state $z_{1}$.}
\label{fig:V030first}
\end{centering}
\end{figure}

In support of our statement that there is interference between the family of
long-lived resonance states and the type II Stark resonances we include
Figs.(\ref{fig:V030}) and (\ref{fig:V030first}). In these Figures, we plot
the absolute value of the wavefunction $z_{1}$ and $z_6$ shown in Table III, as a function of the one-dimension $x$
for $V_{0}=30$. Using Eq.(\ref{eq:squareresonances}) one
finds that the lowest type II Stark resonance of the field-free square well is
$E_{R,4}=9.478$, for $V_{0}=30$. The Stark resonance $z_{6}$ in Table III
stems from the $E_{R,4}$ state. In
Fig.(\ref{fig:V030}) we see that $|\Psi_{z_{6}}(x)|$ has a nodal
structure with four pronounced maxima in the region where the square well extends,
 since it stems from the $E_{R,4}$ resonance state of the field-free square
well. In addition, the nodal pattern of $|\Psi_{z_{6}}(x)|$ involves $6$ maxima localized
in the triangular region above the square well indicating interference with
the family of long-lived states. In
Fig.(\ref{fig:V030first}) we plot $|\Psi_{z_{1}}(x)|$ and we see that
it has one pronounced maximum inside the triangular region above the
square well as well as four less pronounced ones for $-\alpha<x<\alpha$ due to the interference with
the $z_{6}$ Stark resonance.

\subsection{Short-lived ``new-born'' potential-barrier resonance states}

The square well potential plus static electric field supports in
addition to the resonances studied in the previous sections, a family
of an infinite number of short-lived resonance states that are stemming from
the continuum. For $V_{0}=5$, these resonance states are denoted as $z_{-1},
z_{-2},...$ in Table I. These short-lived states are overlapping
resonances, that is, the spacing of the real energy part is less than the
width of these states. When $V_{0}\rightarrow \infty$ one can show that the condition
for obtaining these states is $Ci(-z_{n}-\alpha)=0$ \cite{Ludvikson}. That is,
 to the lowest order in 1/V, these short-lived states are given by
 $(-3\pi/8+3l\pi/2)^{2/3}e^{-i 2\pi/3}-\alpha$, with $l=1,2,...$.

Understanding the ``nature" of these broad resonances is harder than
understanding the ``nature" of the long-lived ones for the following reasons.
 First, in the limit $V_{0}$ becomes very
 large, the real part of the long-lived states are the eigenenergies of the triangular well formed by the static
 electric field for $x<-\alpha$ and a wall at $x=-\alpha$, as discussed in
section IVA, see also ref \cite{Agapi} and references there in. The fact that
the long-lived ``new-born'' states are localized above the square well and
primarily in the above mentioned triangular region is also corroborated by
the nodal structure of these states shown in Fig.(\ref{fig:absvalue}).  
 It is
 not known, however, what part of the potential are the
 short-lived states associated with. The fact that these resonance
states do not have a nodal structure, see Fig.(\ref{fig:absvalue}),
 does not make it evident what part of the potential these states are 
associated with.  
 Second, in a 1D scattering experiment, that
 one measures, for example, the wigner delay
time as a function of the incident energy one finds {\it peaks}
for energies equal to the real energy of the long-lived states but
one does not find a {\it peak structure} for the broad overlapping
resonances \cite{Agapi}. In photodetachment experiments of
negative ions driven by a week time periodic field in the presence
of a static electric field \cite{Hion1,Hion2}, the long-lived
states are known to give rise to a {\it peak structure}
\cite{Hion1,Hion2} while the short-lived ones do not. These latter
ones however have been shown to have observable effects in the
photodetachment rate of the negative ions when driven by a weak
time periodic field in the presence of a static field
\cite{Agapi}. Mainly, the short-lived resonance states stemming
from the continuum of the $\delta$-potential plus static electric
field \cite{Ludvikson} have been shown to be responsible for the
asymmetry  of the {\it peaks} as well as the considerable
photodetachment rate for photon energies smaller than the binding energy
of the loosely bound electron of the $H^{-}$ ion, that is experimentally observed
as a ``shoulder" structure.

 It is thus of interest to elucidate the ``nature'' of the short-lived
resonances and suggest a possible experimental set up where they
could give rise to {\it peaks} as is the case for the long-lived
``new-born'' resonances. To do so, we first show that the
short-lived states are associated {\it only with the static
electric field potential barrier}. Then, by introducing a cut-off
and thus truncating the static electric field, we find broad
overlapping resonances, that are again associated {\it only with
the static electric field potential barrier}. Using complex
scaling, we show that {\it the nodal structure is hidden under the
natural exponential divergent nature of these broad resonance
states}. Lastly, we discuss a 2D scattering set-up where the broad
resonances of the square well plus static field should in
principle give rise to {\it peaks}.

\subsubsection{Potential barrier resonances}

\begin{figure}
\begin{centering}
\leavevmode
\epsfxsize=0.6\linewidth
\epsfbox{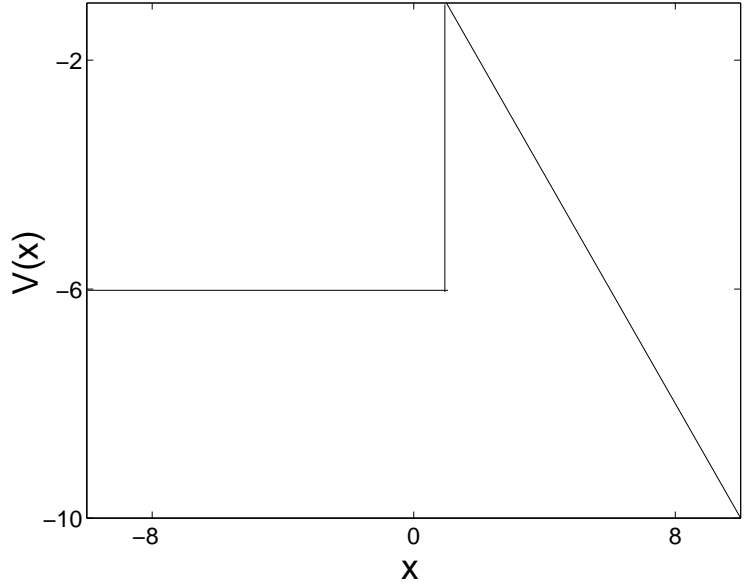}
\caption{The potential barrier}
\label{fig:f6}
\end{centering}
\end{figure}

In what follows, we show that the short-lived overlapping
resonance states stemming from the continuum are {\it associated
only with the potential barrier formed by the static electric
field for $x>\alpha$}, see Fig.(\ref{fig:f6}).
 We take the potential barrier to be
\begin{equation}
\label{eq:Vbarrier}
V(x)=\left\{\begin{array}{ll}
 -V_{0}-\alpha,& \mbox{$x< \alpha$}\\
  -x, &\mbox{$x>\alpha$}
                             \end{array}
                              \right.
\end{equation}
To find the resonance states of the potential barrier we follow the
same approach as in section II, that is, we find the wavefunction
satisfying outgoing wave boundary conditions which is found to be:
\begin{equation}
\label{eq:continuumb}
\Psi(x,z)=\left\{\begin{array}{ll}
 N e^{-ikx}, & \mbox{$x< \alpha$} \\
 N b_{1}(z)Ci^{+}(-x-z),& \mbox{$x> \alpha$}\\
                             \end{array}
                              \right.
\end{equation}
where $k=\sqrt{V_{0}+\alpha+z}$.
 From the continuity of the wavefunction and its first derivative at $x=\alpha$
we obtain the condition for resonance states:
\begin{equation}
\label{condresonances1}
Ci^{+}(-\alpha-z_{n})ik-Ci'(-\alpha-z_{n})=0
\end{equation}
where $Ci'(-x-z_{n})$ is the derivative with respect to x.
  We choose the
$Im(k)>0$ so that $\Psi(x,z)\rightarrow 0$ as $x\rightarrow
 -\infty$. As $x\rightarrow \infty$, $\Psi(x,z_{n}) \rightarrow \infty$.
 We find that the potential barrier, supports a family of an infinite number
  of short-lived resonance states. {\it These states are given in Table IV,
 and a comparison with Table I reveals that they are almost identical with the
short-lived states of the square well plus static electric field.}

The one-dimensional symmetric and nonsymmetric Eckart potential
barrier are smooth barriers that support short-lived overlapping
states, as has been shown by Ryaboy and Moiseyev \cite{Ryaboy}. In
ref\cite{Ryaboy}, it was shown that after complex scaling, the
wavefunctions of these short-lived states are localized inside the
potential barrier and have a nodal structure. It is also
interesting to note that in contrast to the sharp edges of the
potential barrier currently under consideration the Eckart barrier
is a smooth potential. We believe that this latter fact suggests
that the existence of the short-lived resonance states of the
square well plus static electric field is not due to the sharp
edges of this potential.

 So far, we have shown, that the short-lived overlapping
resonance states of the one-dimensional square well plus static
electric field are {\it eigenstates of the static electric field potential
barrier}, as is the case for the Eckart potential-barrier resonances.
 However, unlike the Eckart potential-barrier resonances the short-lived
overlapping resonances of the square well plus static electric field do not
have a nodal structure, as we have shown. In the next section, we show that
by truncating the static electric field one finds a nodal structure for
overlapping resonances associated only with the static field potential barrier.

\subsubsection{Square well plus static electric field with cut-off}

In what follows, we truncate the static electric field by
introducing a cut-off. We find that the square well plus the
truncated static electric field supports, in addition to other
resonances, broad overlapping ones which are again associated {\it
only with the static electric field potential barrier} and they
are thus similar in nature with the broad overlapping resonances
of the square well plus static electric field. Using, complex
scaling we show that the overlapping resonances that are
associated with the truncated potential barrier have a nodal
structure. This suggests that it can be the case that the broad
overlapping resonances can have observable effects in the
transmission properties of this 1D truncated potential. It would
be of interest in the future to study the effect of the potential
barrier resonances on the transmission properties of the truncated
1D system with $N>1$ quantum wells.

 The one dimension potential is given by:
\begin{equation}
\label{eq:Vsquarecut}
V(x)=\left\{\begin{array}{llll}
 -x, & \mbox{$x< -\alpha$} \\
 -V_{0}-x,& \mbox{$-\alpha<x< \alpha$}\\
  -x, &\mbox{$\alpha<x<x_{0}$}\\
  -x_{0},&\mbox{$x>x_{0}$}

                             \end{array}
                              \right.
\end{equation}
while the wavefunction, after applying outgoing boundary conditions,
 is given by:

\begin{equation}
\label{eq:continuumc}
\Psi(x,z_{n})=\left\{\begin{array}{llll}
 N Ai(-x-z_{n}), & \mbox{$x< -\alpha$} \\
 N (c_{1}Ai(-x-V_{0}-z_{n})+c_{2}Bi(-x-V_{0}-z_{n})),& \mbox{$-\alpha<x<
 \alpha$}\\
 N(c_{3}Ai(-x-z_{n})+c_{4}Bi(-x-z_{n})), & \mbox{$\alpha<x<x_{0}$}\\
 Nc_{5}e^{ikx},& \mbox{$x>x_{0}$}\
                             \end{array}
                              \right.
\end{equation}
where $k=\sqrt{x_{0}+z_{n}}$. Note that $\Psi(x,z_{n})\rightarrow 0$ as
$x\rightarrow -\infty$, while for $Im(k)<0$ as $x\rightarrow \infty$
$\Psi(x,z_{n})\rightarrow \infty$. The
resonance states of the square well plus static electric field with a cut-off are shown
in Table V for $V_{0}=5$, $\alpha=1$ and $x_{0}=20$.

In the same way as in the previous section, one can show that the
resonance states denoted by b in Table V are almost {\it identical to
the resonance states supported by the potential barrier formed by
the static electric field with a cut-off}.
 By introducing a cut-off the lifetime
of the short-lived states of the square well plus static electric
field increases but they are still {\it overlapping resonances},
see states $z_{-1},z_{-2},...$  in Table V.
 We next apply exterior complex scaling
 to the wavefunction in
Eq.(\ref{eq:continuumc}) to obtain:

\begin{equation}
\label{eq:continuumcc}
\Psi(x,z_{n})=\left\{\begin{array}{llll}
 N Ai(-x-z_{n}), & \mbox{$x< -\alpha$} \\
 N (c_{1}Ai(-x-V_{0}-z_{n})+c_{2}Bi(-x-V_{0}-z_{n})),& \mbox{$-\alpha<x<
 \alpha$}\\
 N(c_{3}Ai(-x-z_{n})+c_{4}Bi(-x-z_{n})), & \mbox{$\alpha<x<x_{0}$}\\
 Nc_{5}e^{ik((x-x_{0})e^{i \theta}+x_{0}))},& \mbox{$x>x_{0}$}\
                             \end{array}
                              \right.
\end{equation}
where $k=\sqrt{x_{0}+z_{n}}$. The normalization is chosen so that
$\int_{-\infty}^{x_{0}}\Psi(x,z_{n})^2 dx +e^{i \theta}\int_{x_{0}}^{\infty}
e^{2ik((x-x_{0})e^{i\theta}+x_{0})} dx =1$.
 Note that $\Psi(x,z_{n})\rightarrow 0$ as $x\rightarrow \infty$
and in addition the wavefunctions of different resonance states
are orthogonal, that is, $\int_{-\infty}^{+x_{0}}\Psi(x,z_{n})
\Psi(x,z_{n'}) dx + e^{i
\theta}\int_{x_{0}}^{\infty}\Psi(x,z_{n})\Psi(x,z_{n'})
dx=\delta_{n,n'}$. $\theta$ is chosen so that $\theta>\phi$ with
$x_{0}+z_{n}=k^2=|k|^2e^{2i\phi}$. In
Fig.(\ref{fig:barrier5firstneg})  we show that the complex scaled
wavefunctions of the overlapping resonances $z_{-1},z_{-2},...$
have a {\it nodal structure which is explored due to the use of
the exterior scaling transformation}.
 For an explanation of the complex scaling transformations
including the reason we do not take the complex conjugate of the
``bra" state when we calculate expectation values, as is usually
done when the conventional scalar product is applied, see
Ref.\cite{Moiseyev}.

\begin{figure}
\begin{centering}
\leavevmode
\epsfxsize=0.6\linewidth
\epsfbox{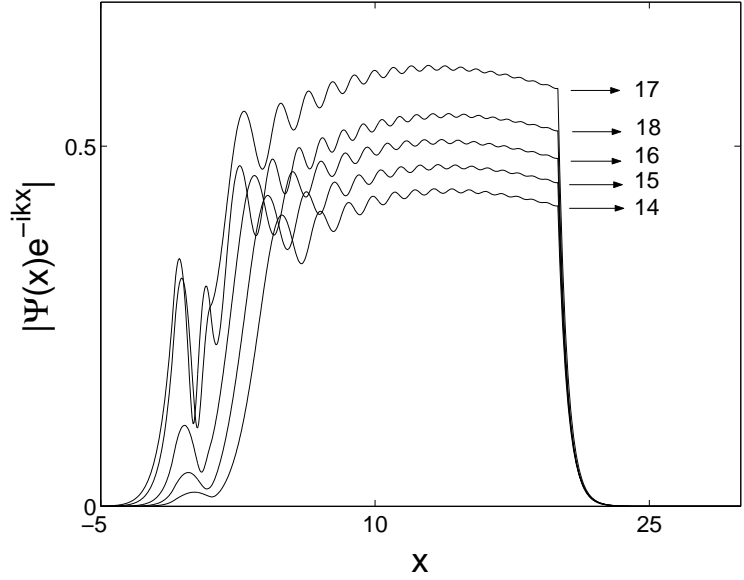}
\caption{$|\Psi_{n}(x)e^{-ikx}|$ for the $z_{-1},z_{-2},z_{-3},z_{-4},z_{-5}$
states in Table V. Their nodal pattern consists of 18,17,16,15,14 maxima
respectively. This nodal pattern is localized inside the
region defined by the static electric field for $x>\alpha$ and a wall at $x=x_{0}$.}
\label{fig:barrier5neg}
\end{centering}
\end{figure}

\begin{figure}
\begin{centering}
\leavevmode
\epsfxsize=0.6\linewidth
\epsfbox{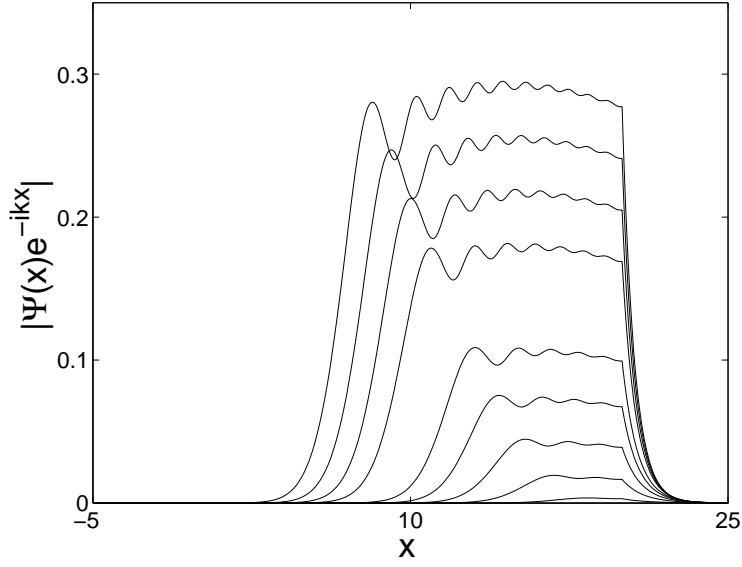}
\caption{$|\Psi_{n}(x)e^{-ikx}|$ for the
$z_{-17},z_{-16},...,z_{-9}$
states in
Table V with
a nodal pattern of 1,2,...,10 maxima respectively. This nodal pattern is localized inside the
region defined by the static electric field for $x>\alpha$ and a wall at $x=x_{0}$.}
\label{fig:barrier5firstneg}
\end{centering}
\end{figure}

\begin{figure}
\begin{centering}
\leavevmode
\epsfxsize=0.6\linewidth
\epsfbox{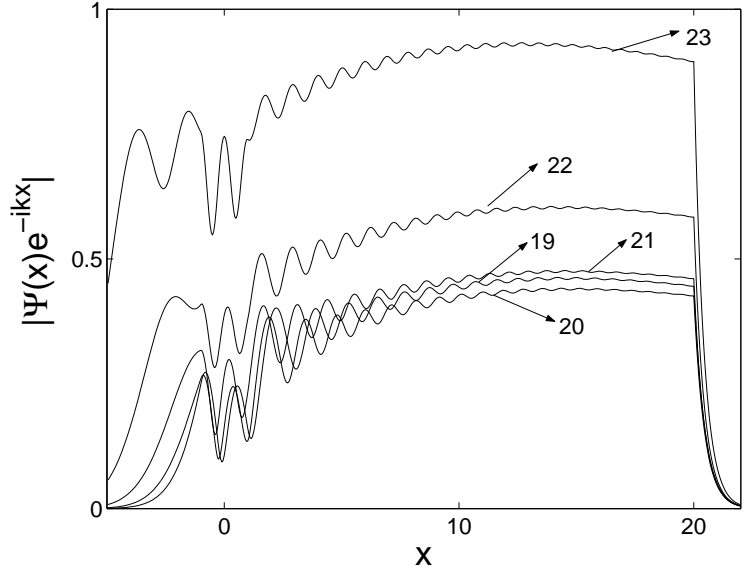}
\caption{$|\Psi_{n}(x)e^{-ikx}|$ for the $z_{2},z_{3},z_{4},z_{6},z_{8}$ in
Table V with a nodal pattern of 19,20,21,22,23 maxima respectively. This nodal
pattern is localized in the region between
$\alpha$ and $x_{0}$.}
\label{fig:barrier5higher}
\end{centering}
\end{figure}

As it is shown in Fig.(\ref{fig:barrier5firstneg}), the complex
scaled wavefunction of the states $z_{-17}, z_{-16},..., z_{-9}$
have a nodal structure with 1,2,...,10 maxima respectively.  In
Fig.(\ref{fig:barrier5neg}) we can see the nodal structure of the
states $z_{-1},z_{-2},z_{-3},z_{-4},z_{-5}$ with 18,17,16,15,14
maxima respectively. This nodal pattern, if $|\Psi(x)e^{-ikx})|$
is shifted by the real part of the resonance, is localized in the
region formed by the static electric field for $x>\alpha$ and a
wall at $x=x_{0}$. Therefore, the states are not localized inside
the potential barrier as in the Eckart potential case but {\it
over} the barrier. Note, however, that these resonances are still
considered as {\it potential-barrier resonances since they ``pop
up'' only when the potential barrier appears and even in the
absence of the field-free square potential well}. In
Fig.(\ref{fig:barrier5higher}) we plot the states
$z_{2b},z_{3b},z_{4b},z_{6b},z_{8b}$ which also have a nodal
pattern with 19,20,21,22,23 maxima and when shifted by the real
part of the energy seem to be localized between $\alpha$ and
$x_{0}$ above the well. 

We have thus shown in this section that the states indicated by b
in Table V are eigenstates of the potential barrier with a
cut-off, are localized over the barrier, and definitely not inside
the potential well, and have a nodal structure.

\begin{figure}
\begin{centering}
\leavevmode
\epsfxsize=0.6\linewidth
\epsfbox{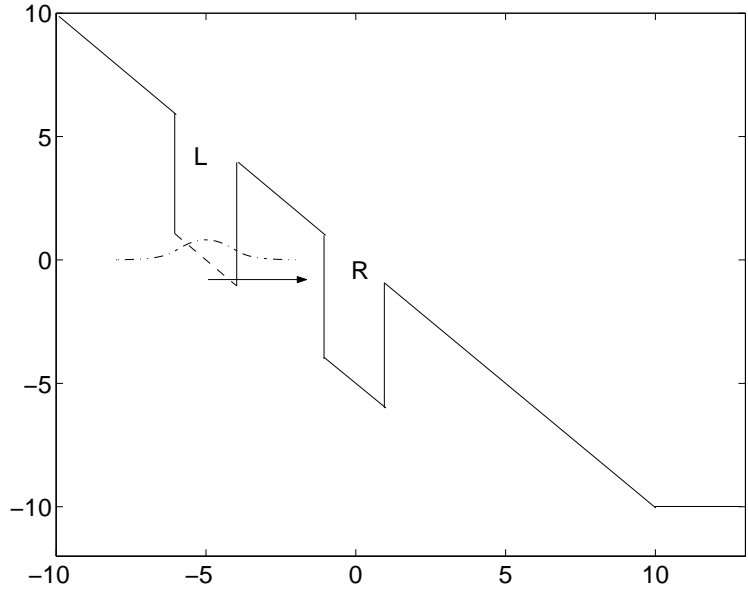}
\caption{The wavefunction initially localized at the left well is tunnelling
to the region with $x>x_{0}$ where $x_{0}$ is taken equal to $10$ in this plot.}
\label{fig:2Dwells}
\end{centering}
\end{figure}

As we have already mentioned, it might be interesting to study in the future
 the effect of the
long-lived as well as of the overlapping ``new born" resonances in a 1D scattering
experiment with $N>1$ quantum wells in the presence of a static field with a
cut-off. If for example one considers two quantum wells plus a static field
with a cut-off, see Fig.(\ref{fig:2Dwells}), one could consider an initial
wavefunction localized in one of the quantum wells and compute the probability
for this wavefunction to tunnel to the region with constant potential, that
is, to the region with $x>x_{0}$. The 
long-lived ``new born'' resonances should be observable as ``peaks'' in the
transmission probability when plotted as a function of energy but it also
might be the case that the overlapping ``new born'' resonances affect, in some
observable way, the transmission probability as well.

\subsection{2D experimental setup}

Finally, we propose a two-dimensional (2D) experimental set up
where the short-lived overlapping resonances of the square well
plus static electric field (without a cut-off), that were studied in section
IVB1, should in principle give rise to a {\it peak
structure}. Our proposed 2D set-up is based on an idea by
Narevicius and Moiseyev \cite{Narevicius}. In ref\cite{Narevicius}
the short-lived resonance states of the 1D
Eckart potential \cite{Ryaboy} which are localized in a
classically inaccessible region have been considered. It has been
shown that these resonances become a very good approximation for
2D resonance states that do yield a {\it peak structure} and are
localized in a classically accessible region.
 That idea is based on the trapping of a light atom between two heavy
ones. The motion of the light atom along the coordinate $x$ perpendicular
to the distance $y(x)$ between the heavy atoms is treated as the slow one.
 This is a reasonable approximation if it is for
the light particle to get temporarily trapped between the heavy atoms.
 In this adiabatic approximation the eigenenergies along the $y(x)$
direction become the adiabatic potentials along the $x$ direction, which in the case
of ref\cite{Narevicius} is the 1D Eckart potential.

\begin{figure}
\begin{centering}
\leavevmode
\epsfxsize=0.6\linewidth
\epsfbox{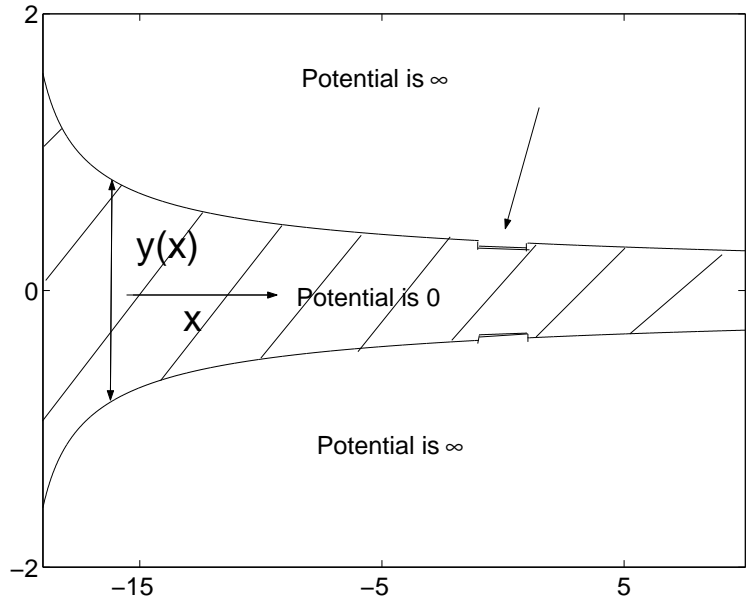}
\caption{2D experimental set up}
\label{fig:experiment}
\end{centering}
\end{figure}

 Given the above idea we propose the following 2D potential to observe the
short lived resonance states of the square well plus static electric
field: The electrons move ``freely'' in a 2D quantum well/dot that has the shape shown in Fig.(\ref{fig:experiment}). This 2D potential is associated with
a $y(x)$ which is given by,
\begin{equation}
\label{eq:Vsquarecut2}
y(x)=\left\{\begin{array}{llll}
  \pi/2\times 1/\sqrt{x+\Delta_{0}},&\mbox{$-\Delta_{0}<x<-\alpha$}\\
 \pi/2\times 1/\sqrt{+x+\Delta_{0}+V_{0}}, & \mbox{$-\alpha<x< \alpha$} \\
 \pi/2\times 1/\sqrt{+x+\Delta_{0}},& \mbox{$x> \alpha.$}

                             \end{array}
                              \right.
\end{equation}
The motion of the electrons along the $y(x)$ direction is much faster than the
motion along the $x$ direction. Therefore the adiabatic approach is
applicable here as in the case studied before when a light atom is scattered
from two heavy ones. The eigenenergy of the light particle along the $y(x)$ direction is the
ground state of a particle in a box. This eigenenergy is also the
adiabatic potential along the $x$ direction which we take it to be the
square well plus static electric field.
 For a particle in a box the ground state energy is given by
\begin{equation}
\label{energy}
E_{0}=-\pi^2 /(4y(x)^{2}).
\end{equation}
We design our proposed experiment such that $y(x)$ is given by
$y(x)=\pi/(2\times \sqrt{|E_{0}|})$ with $E_{0}$ being the ground
state of the square well which in our case is given by the
potential in Eq.(\ref{eq:potential1}) shifted downwards by
$\Delta_{0}$. The downward shift introduces a cut-off for
$x<-\Delta_{0}$ such that on one hand it does not affect the short-lived
states and on the
other it allows for a finite potential for
$x<-\Delta_{0}$.

\section{Conclusions}
In conclusion, using a simple model, a square well in the presence
of a static electric field we have shown that, besides the Stark resonances,
 there are two-families of resonances that are not associated with
the spectrum of the field-free potential. One family consists of
long-lived states that are localized above the square well. The
other family consists of short-lived overlapping resonances that
in a 1D scattering experiment do not give rise to a {\it peak structure}. We have shown
that these short-lived resonance states are associated with the
potential barrier. By introducing a cut-off in the square well
plus static electric field we have found again overlapping
resonances which are associated only with the static field
potential barrier and that have nodal structure. Future studies
could involve studying the effects of these potential barrier
resonances on the transmission properties of the truncated 1D
potential with $N>1$ wells. Finally, we have proposed a 2D
scattering experiment where in principle the short-lived resonance
states of the square well plus static electric field should yield
a {\it peak structure}.

\newpage
\begin{center}
Tables
\end{center}

\begin{center}
Table I: Poles $z_{n}$ of the square well plus static electric field in the energy
range $-5<Re(z_{n})<12$, for $V_{0}=5$.
\end{center}
\vspace {1cm}
\begin{center}
\begin{tabular}{ccc} \hline
n & $z_{n}$  &
 \\ \hline
 -5 &  -4.8582935-7.3068739*i \\
 -4  & -4.2903538-6.3150215*i \\
 -3 &  -3.6721261-5.2295407*i \\
 -2 & -2.9761088-3.9997772*i  \\
 -1  & -2.1214686-2.5051412*i \\
  0  & -3.970613-0.0004836*i  \\
  1  & -1.0511476-0.3388278*i  \\
  2  & 3.0676534-0.3872741*i \\
  3  & 4.5571629-0.7205034*i   \\
  4  & 5.4849666-0.5994673*i \\
  5 &  6.7267609-0.3737202*i\\
  6 &  7.9088730-0.2989037*i\\
  7 &  9.0068094-0.2684895*i\\
  8 &  10.0385628-0.2575935*i\\
  9 &  11.0173509-0.2578512*i\\
  10 &  11.9525012-0.2658891*i\\ \hline
\end{tabular}
\end{center}

\begin{center}
Table II: Eigenenergies of the triangular well formed by the static electric field in
the region $x<-\alpha$ and a wall at $x=-\alpha$.
\end{center}
\begin{center}
\begin{tabular}{ccc} \hline
l & $E$  &
 \\ \hline
 1 & 3.32025\\
 2  & 5.08181\\
 3 &  6.51716\\
 4 &  7.78445\\
 5 &  8.94249\\
 6 & 10.0214\\
 7 & 11.0391\\
 8 & 12.0077\\ \hline

\end{tabular}
\end{center}

\begin{center}
Table III: Poles $z_{n}$ of the square well plus static electric field in the energy
range
 $0<Re(z_{n})<10$ for $V_{0}=30$.
\end{center}
\begin{center}
\begin{tabular}{ccc} \hline
n & $z_{n}$  &
 \\ \hline
 1 & 3.2499347-0.0765441*i\\
 2&  4.9503494-0.1167936*i\\
 3 & 6.3426275-0.1714228*i\\
 4 & 7.5729956-0.2513833*i\\
 5 &8.6823136-0.3926178*i\\
6&  9.4906152-0.57571047*i\\ \hline
\end{tabular}
\end{center}

\begin{center}
Table IV: Poles $z_{n}$ of the barrier shown in Fig. 4 in the
energy range $-5<Re(z_{n})<0$ . \vspace {1cm}
\begin{tabular}{ccc} \hline
n & $z_{n}$  &
 \\ \hline
 -5 &   -4.8574370-7.2999738*i\\
 -4  & -4.2885652-6.3071804*i \\
 -3 &  -3.6688695-5.2205217*i \\
 -2 & -2.9713127-3.9888788*i  \\
 -1  & -2.1234444-2.47845124*i \\ \hline
\end{tabular}
\end{center}

\begin{center}
Table V: Poles $z_{n}$ of the square well plus static electric field with cut-off for
$-20<Re(z_{n})<7$ with $x_{0}=20$ and $V_{0}=5$.
\end{center}
\begin{center}
\vspace {1cm}
\begin{tabular}{ccc} \hline
n & $z_{n}$  &
 \\ \hline
 -17b  &  -17.2097016-1.0943095*i\\
 -16b  &  -15.5425549-1.0265055*i\\
 -15b  &  -14.1537578-0.9791738*i\\
 -14b  &  -12.9162592-0.9431633*i\\
 -13b  &  -11.7796093-0.9142266*i\\
 -12b  &   -9.7125118-0.8694805*i\\
 -11b  &   -8.1548697-0.8514889*i\\
 -10b  &   -7.8364663-0.8355591*i\\
 -9 b  &   -6.9515568-0.8212841*i\\
 -8 b  &   -6.0957260-0.8083655*i\\
 -7 b  &   -5.2654870-0.7965786*i\\
 -6 b  &   -4.4580053-0.7857675*i \\
 -5 b &  -3.6708589-0.7761792*i  \\
 -4 b &   -2.9016305-0.7689446*i \\
 -3 b &   -2.1417117-0.7637019*i  \\
 -2 b &  -1.3432008-0.7307476*i \\
 -1 b   &  -0.40709318-0.8295943*i \\
  0   &   -3.97061424-0.0004822*i\\
  1   &   -1.0386799-0.3352285*i\\
  2 b  &   0.4649898-0.9253466*i\\
  3 b  &   1.3855199-0.9779873*i\\
  4 b  &   2.3564670-0.9696458*i \\
  5   &  3.0590434-0.3809258*i\\
  6 b  &  3.4826672-0.9145274*i\\
  7   &  4.4107251-0.6948808*i\\
  8 b  &  4.8838984-0.7687683*i\\
  9   &  5.5023958-0.5534036*i\\
  10 b &  6.2087394-0.9667513*i\\
  11  &  6.7318740-0.3683771*i\\ \hline
\end{tabular}
\end{center}

\clearpage


\end{document}